\documentclass[twocolumn,nopacs,preprintnumbers,amsmath,amssymb,subfig]{revtex4}

\usepackage{color}
\usepackage{graphicx}
\usepackage{multirow}
\usepackage{comment}

\begin{document}

\title{Magnetic field-induced degenerate ground state in the classical antiferromagnetic XX model on the icosahedron}

\author{N. P. Konstantinidis}
\affiliation{Mathematics and Science Department, American University in Bulgaria, Sv. Buchvarova Str. 8, Blagoevgrad 2700, Bulgaria}

\date{\today}

\begin{abstract}
The ground state of the classical antiferromagnetic XX model in a magnetic field is calculated for spins mounted on the vertices of the icosahedron. The magnetization is characterized by two discontinuities as a function of the external field. For a wide field range above the first discontinuity the ground state is degenerate, with two spins related by spatial inversion almost aligned with the field and the rest forming two magnetization units in the form of pentagons. It is shown that the degeneracy originates from the coupling of the two pentagons, which introduces the triangle, associated with ground-state degeneracy, as an interaction unit in the icosahedron. The magnetization discontinuities are shown to evolve first from the coupling of isolated triangles and then from the addition of the two spins related by spatial inversion.
\end{abstract}


\maketitle

\section{Introduction}
\label{sec:introduction}

The description of strong electron correlations is often reduced to interactions between localized electrons via their spin degrees of freedom \cite{Auerbach98,Fazekas99}. Such interactions have been investigated on various lattices and molecules with a single spin mounted on each vertex. The topology of frustrated structures introduces competing interactions, where not all interacting spin pairs can simultaneously minimize their energy in the classical ground state for isotropic antiferromagnetic interactions \cite{Manousakis91,Lhuillier01,Misguich03,Ramirez05,Schnack10}.  The spins are typically taken to be three-dimensional, however classical spins defined in more than three dimensions have also been considered for a variety of molecular structures, with their interactions described by the $n$-vector model \cite{NPK23}. Spins existing in more than three dimensions allow for further minimization of the interaction energy in the case of frustration, and the number of dimensions where the lowest-energy spin configuration is found is a criterion for the strength of frustration of the underlying structure.

The addition of a magnetic field introduces a new interaction scale, resulting in competition between the frustrated spin interactions and the field energy for minimization. This can produce a magnetic response that is far from smooth, with multiple magnetization and susceptibility discontinuities for both classical and quantum spins interacting according to the antiferromagnetic Heisenberg model (AHM) \cite{Coffey92,NPK05,NPK07,NPK16,NPK23-1,NPK25,NPK17,NPK23-2,NPK21,Schulenburg02,Richter04,Schnack06,Nakano13,Nakano14,Nakano14-1,Furuchi21,Furuchi22,Schmidt24,Richter25}. The icosahedron, a Platonic solid \cite{Plato}, has been found to achieve its zero-field absolute ground state in three spin dimensions within the framework of the $n$-vector model \cite{NPK23}. In this paper, in order to examine the consequences of dimensional confinement and the nature of coplanar ground states, the spins are restricted in two dimensions and reside on the vertices of the icosahedron. The icosahedron has been found to have a classical magnetization discontinuity for the AHM in a field \cite{Schroeder05}. The magnetic response of a ring of icosahedra has also been calculated for the AHM \cite{NPK25-1}. The molecule has also been considered in different contexts \cite{Vaknin14,Engelhardt14,Hucht11,Sahoo12-1,Strecka15,Karlova16,Karlova16-1,NPK16-1,Suzuki21,Eto25,Watanabe25}. The spins are taken to be classical and to interact according to the antiferromagnetic XX model (AXXM) in an external field. It has been shown that in the zero-field ground state of this model spatial symmetry is lost, with the nearest-neighbor correlations assuming more than one value \cite{NPK23}, unlike the case of three-dimensional spins interacting according to the AHM.

The confinement of the spins in two dimensions results in a total of two magnetization discontinuities in the ground state when the field is ramped up. The first discontinuity eventually drives two spins related by spatial inversion to almost align with the field. These spins are symmetrically oriented with respect to field, and their total magnetization points along it. This results in a degenerate ground state for the icosahedron. Each of the spin-inversion related spins interacts with five spins that form a pentagon. These two pentagons emerge as magnetization units, where the requirements are that their total spin and interaction energies are the same in the ground state. These conditions can be satisfied for a manifold of degenerate ground states, and for each one each pentagon spin typically points in a different direction, simultaneously violating the symmetry of the polar angles with respect to reflections along the $x$ axis. It is shown that the origin of this degeneracy is the coupling of the two pentagons with one another, since then the triangle emerges as an interaction unit. The degenerate ground states of the AXXM on an isolated triangle transfer their degeneracy to the ground state of the
icosahedron for the field window where the two spins related by spatial spin inversion are almost aligned with the magnetic field. Such ground-state degeneracies have been associated with classical spin liquids \cite{Chalker17,Capponi25}.

The low-field magnetization discontinuity is traced to the coupling of the two spins related by spatial inversion to the rest of the icosahedron, as was done for the classical discontinuity of the AHM \cite{NPK15}. In order to locate the origin of the higher-field jump a further dismantling of the icosahedron into individual triangles is required and shows how their coupling produces the discontinuity.





\begin{figure}[h]
\begin{center}
\includegraphics[width=3.2in,height=2.4in]{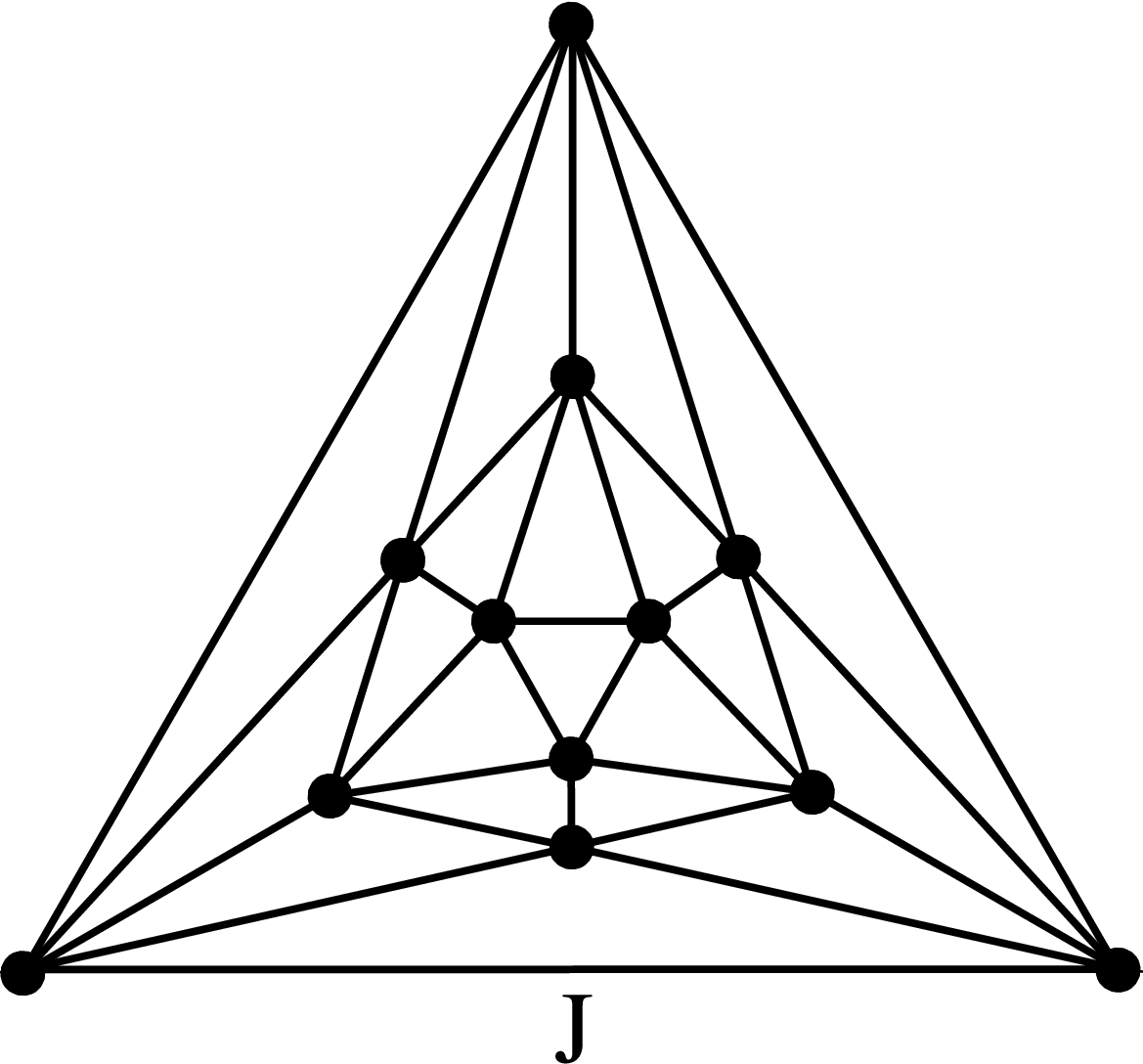}
\end{center}
\caption{A projection of the icosahedron on a plane. The circles are classical spins of unit magnitude and each interacts with its five nearest neighbors with strength $J$, as shown by the connecting lines.
}
\label{fig:icosahedroncluster}
\end{figure}


The plan of this paper is as follows: in Sec. \ref{sec:model} the AXXM in a magnetic field is introduced for the icosahedron, and in Sec. \ref{sec:magnetizationresponse} its ground-state magnetization is calculated. Sec. \ref{sec:connectivityandmagneticresponse} relates the magnetic response to the connectivity of the molecule. Finally Sec. \ref{sec:conclusions} presents the conclusions.

\section{Model}
\label{sec:model}

The icosahedron (Fig. \ref{fig:icosahedroncluster}) has $N=12$ vertices, and on each a classical spin $\vec{s}_i$, $i=1,\dots,N$ of unit magnitude is mounted, defined in two-dimensional spin space. The interactions between spins are described by the Hamiltonian of the AXXM in a magnetic field:

\begin{equation}
H = J \sum_{<ij>} ( s_i^x s_j^x + s_i^y s_j^y ) - h \sum_{i=1}^{N} s_i^x
\label{eqn:model}
\end{equation}

The brackets indicate that the interactions are limited to nearest neighbors $i$ and $j$, which are connected by a molecular edge. $J$ is positive
while the magnetic field points along the $x$ axis. In a bipartite structure nearest-neighbor spins point in antiparallel directions in the zero-field ground state. For frustrated structures the connectivity of the molecule determines the relative orientation of neighboring spins in the absence of a field \cite{NPK23}. The external field in Hamiltonian (\ref{eqn:model}) tends to align the spins along its direction. The magnetic response is determined by the energetic competition between spin interactions and field energy.

The ground-state magnetization response of Hamiltonian (\ref{eqn:model}) was calculated numerically \cite{Coffey92,NPK07,NPK16-1}.
The spins $\vec{s}_i$ are classical unit vectors in two dimensions, and the direction of each is specified by a polar angle ranging from 0 to $2\pi$. The initial spin configuration for a specific field value is selected randomly, and then each angle is moved opposite its gradient direction, until the lowest-energy configuration is reached with double-precision accuracy. To ensure that the ground state is found this procedure is repeated for different random initial configurations.

\section{Magnetization Response}
\label{sec:magnetizationresponse}

The icosahedron consists of 20 triangles and belongs to the icosahedral $I_h$ symmetry point group \cite{Altmann94}. The zero-field ground-state energy of the classical AHM on it equals $-\frac{\sqrt{5}}{5}J$ per interacting pair \cite{Schmidt03}. Its response in a magnetic field is discontinuous \cite{Schroeder05}. Confining the spins in two dimensions results in a zero-field ground state of Hamiltonian (\ref{eqn:model}) lacking the symmetry of the three-dimensional one, with nearest-neighbor spins either antiparallel or forming an angle of $\frac{\pi}{3}$ or $\frac{2\pi}{3}$ \cite{NPK23}.

\begin{figure}[h]
\begin{center}
\includegraphics[width=3.5in,height=2.5in]{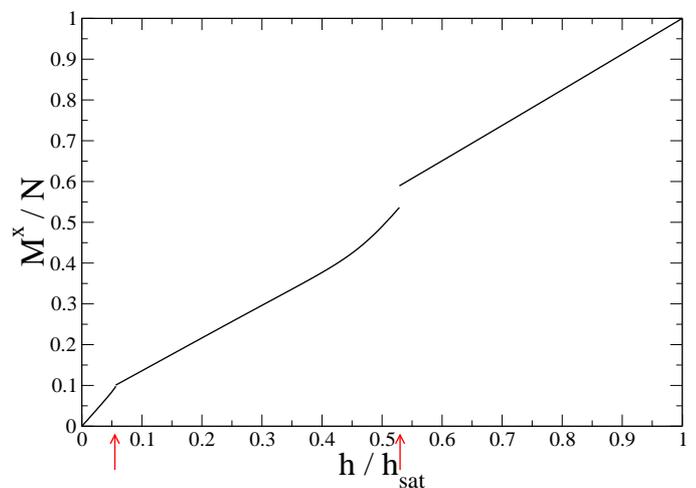}
\end{center}
\caption{Magnetization per spin $\frac{M^x}{N}$ along the field as a function of the magnetic field over its saturation value $\frac{h}{h_{sat}}$ in the ground state of Hamiltonian (\ref{eqn:model}) for the icosahedron. The (red) solid arrows point at the locations of the magnetization discontinuities. The inset focuses on the low-field discontinuity.
}
\label{fig:magnetizationxxicosahedron}
\end{figure}

\begin{figure}[h]
\begin{center}
\includegraphics[width=3.5in,height=2.5in]{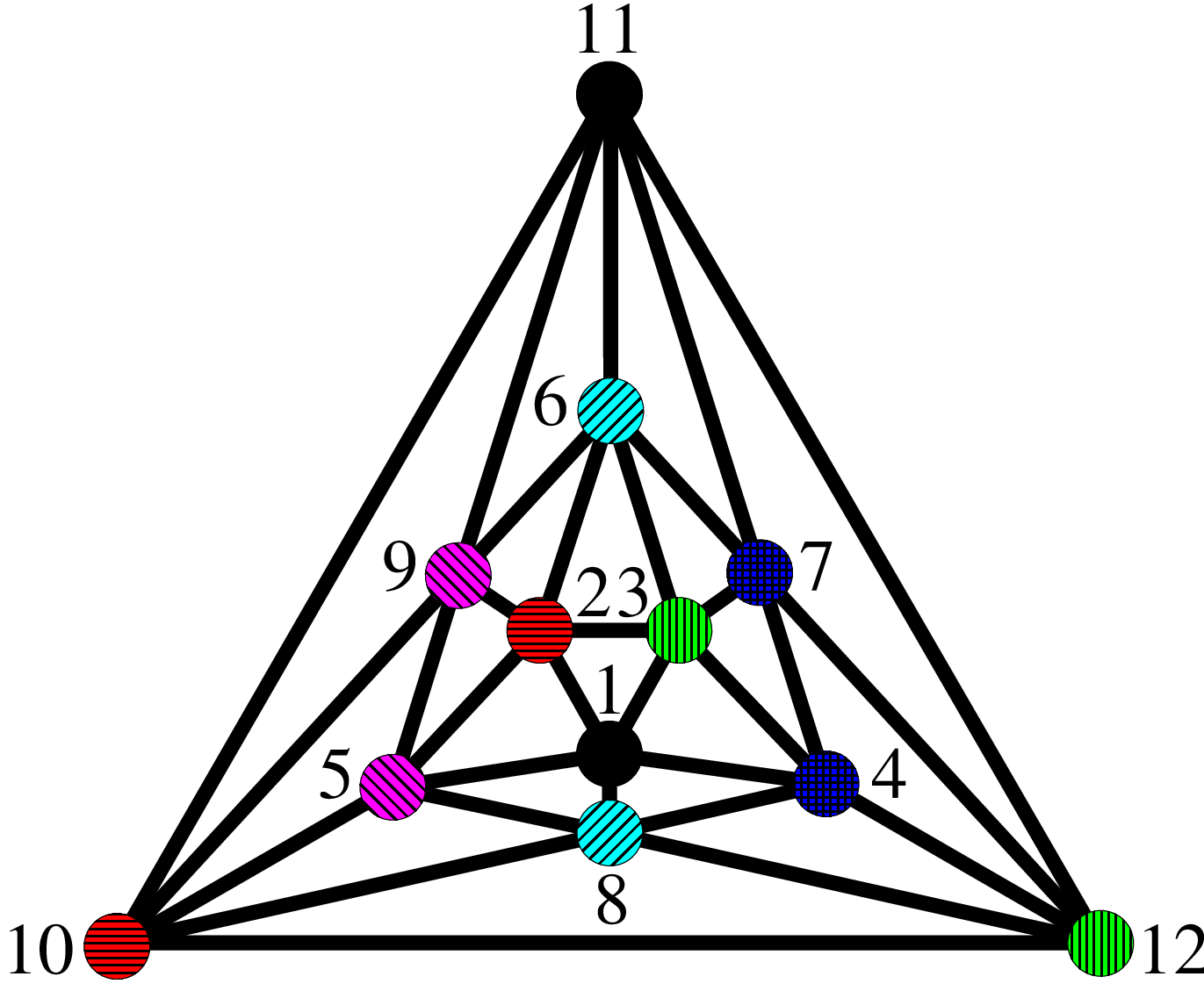}
\end{center}
\caption{Polar-angle configuration of the ground state of Hamiltonian (\ref{eqn:model}) below the first magnetization discontinuity for the icosahedron. Circles with the same pattern (and color) correspond to polar angles adding up to $2\pi$. 
}
\label{fig:icosahedronclusterxx1}
\end{figure}

\begin{figure}[h]
\begin{center}
\includegraphics[width=3.5in,height=2.5in]{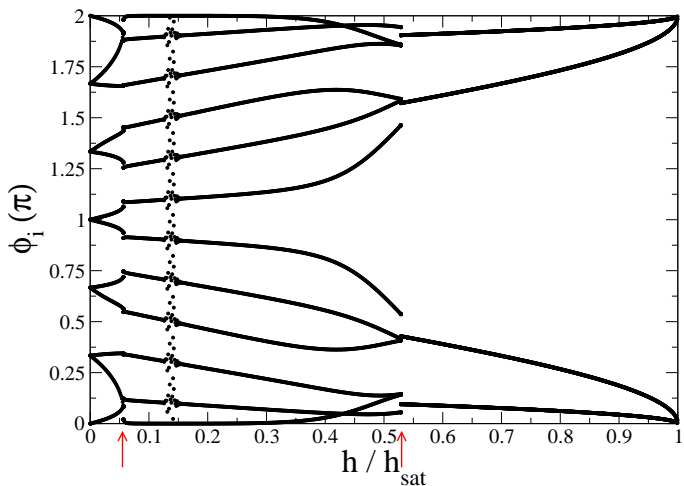}
\end{center}
\caption{Polar angles $\phi_i$, $i=1,\dots,N$ as a function of the magnetic field over its saturation value $\frac{h}{h_{sat}}$ in the ground state of Hamiltonian (\ref{eqn:model}) for the icosahedron. The (red) solid arrows point at the locations of the magnetization discontinuities.
}
\label{fig:polaranglesxxicosahedron}
\end{figure}

The ground-state magnetization of Hamiltonian (\ref{eqn:model}) as a function of the external field is shown in Fig. \ref{fig:magnetizationxxicosahedron}. It is discontinuous at two magnetic field values (Table \ref{table:magnetizationdiscontinuities}). The lowest-energy configuration below the first discontinuity is shown in Fig. \ref{fig:icosahedronclusterxx1} and is characterized by six unique polar angles.
The polar-angle values are plotted in Fig. \ref{fig:polaranglesxxicosahedron} and the nearest-neighbor correlations in Fig. \ref{fig:correlationsxxicosahedron}. The polar angles below the first discontinuity are symmetric with respect to reflections along the $x$ axis.

\begin{table}[h]
\begin{center}
\caption{Magnetization discontinuities of the ground state of Hamiltonian (\ref{eqn:model}). The table lists the magnetic field $h$ of the discontinuity over the saturation field $h_{sat}$, the magnetization per site just below and just above the discontinuity $\frac{M_-^x}{N}$ and $\frac{M_+^x}{N}$, and the change in the magnetization per site $\frac{\Delta M^x}{N}$ in the discontinuity.
}
\begin{tabular}{c|c|c}
$\frac{h}{h_{sat}}$ & 0.0565953756427  & 0.52886352217637 \\
\hline
$\frac{M_-^x}{N}$ & 0.0981229 & 0.53631792245598 \\
\hline
$\frac{M_+^x}{N}$ &  0.1007458 & 0.58935472614845 \\
\hline
$\frac{\Delta M^x}{N}$ & 0.0026229 & 0.05303680369247
\end{tabular}
\label{table:magnetizationdiscontinuities}
\end{center}
\end{table}

\begin{figure}[h]
\begin{center}
\includegraphics[width=3.5in,height=2.5in]{correlationsxxicosahedron}
\end{center}
\caption{Nearest-neighbor correlations $\vec{s}_i \cdot \vec{s}_j$ as a function of the magnetic field over its saturation value $\frac{h}{h_{sat}}$ in the ground state of Hamiltonian (\ref{eqn:model}) for the icosahedron. The (red) solid arrows point at the locations of the magnetization discontinuities.
}
\label{fig:correlationsxxicosahedron}
\end{figure}

\begin{figure}[h]
\begin{center}
\includegraphics[width=3.5in,height=2.8in]{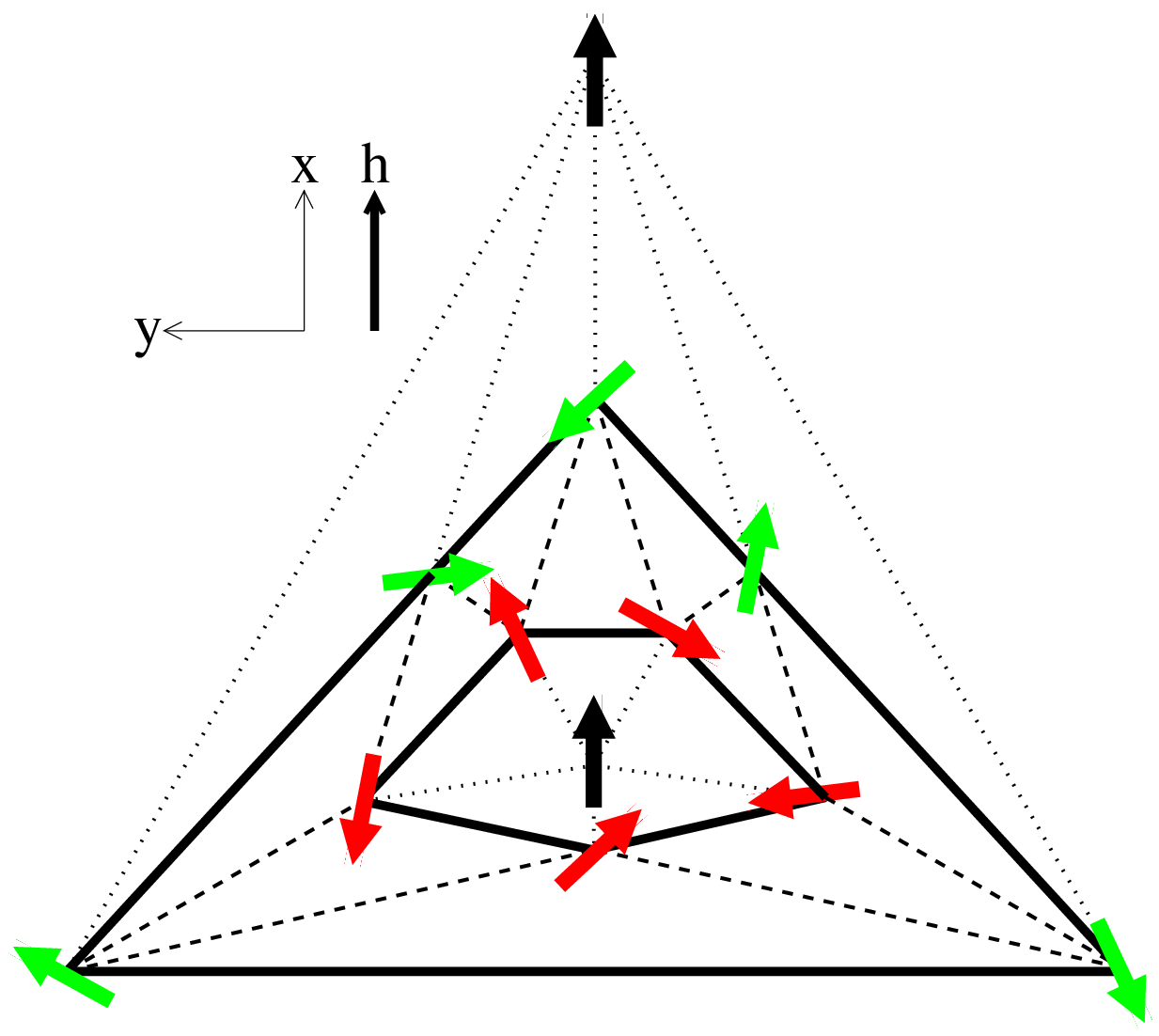}
\end{center}
\caption{The arrows show the spin configuration of the ground state of Hamiltonian (\ref{eqn:model}) for $h=J$ for the icosahedron. The solid lines highlight the two pentagons, each of which consists of nearest neighbors of a spin pointing along the magnetic field direction $\hat{x}$, shown in black. The spins of the two corresponding pentagons are highlighted in red and green. The dashed lines connect spins belonging to different pentagons. The dotted lines connect the two spins aligned with the field with their nearest neighbors. These two spins are related by spatial inversion. The ground-state energy is invariant under simultaneous rotation of all pentagon spins around the $z$ axis, which is perpendicular to the plane of the page, by an angle ranging continuously from 0 to $\frac{\pi}{5}$.
}
\label{fig:icosahedronclusterh=1}
\end{figure}

At the first magnetization discontinuity the symmetry of the ground state does not change, remaining the same it was for fields below the discontinuity, however the polar angles are discontinuous. Following the discontinuity, two spins related by spatial inversion (for example 1 and 11 in Fig. \ref{fig:icosahedronclusterxx1}) become almost parallel to the field, with their total magnetization pointing along it (Fig. \ref{fig:polaranglesxxicosahedron}). At the special value $h=J$, where $\frac{h}{h_{sat}}=\frac{1}{5+\sqrt{5}}$, the ground state is highly degenerate (Fig. \ref{fig:icosahedronclusterh=1}). The two spins related by spatial inversion are now aligned with the field. The configuration of the rest of the cluster is identical to the one it has at zero field when isolated. The nearest-neighbors of each one of the field-aligned spins define two pentagons (formed by spins 2, 3, 4, 5, 8 and 6, 7, 9, 10, 12 in Fig. \ref{fig:icosahedronclusterxx1}). The configuration of each pentagon is the one of the ground state of an isolated pentagon, with the angles between nearest-neighbors equal to $\frac{4\pi}{5}$ and the total magnetization equal to zero. Nearest-neighbor spins belonging to different pentagons form an angle of $\frac{3\pi}{5}$ \cite{Referee25}. Summarily, in the $h=J$ ground state two of the spins are pointing in the direction of the field, while the other ten are evenly distributed in the $xy$ plane, with successive spins forming an angle of $\frac{\pi}{5}$. This is reminiscent of the high-field lowest-energy configuration of the AHM \cite{Schroeder05}. The two maximally polarized spins isolate the rest of the cluster from the magnetic field, allowing it to assume its zero-field ground state. The ground-state energy is invariant under simultaneous rotation of only the ten spins belonging to the two pentagons around the $z$ axis, which is perpendicular to the plane of the page, with their angle of rotation $\omega$ ranging continuously from 0 to $\frac{\pi}{5}$. Even though in such a rotation the angles between each field-aligned spin and its nearest-neighbors change, its total energy with its neighboring pentagon remains the same, and eventually the overall ground-state energy, $-5\sqrt{5}J-2h$ (App. \ref{appendix:h=J}), does not change. What is lost is the reflection symmetry of the spin configuration with respect to the $x$ axis.

\begin{figure}[h]
\begin{center}
\includegraphics[width=3.5in,height=2.5in]{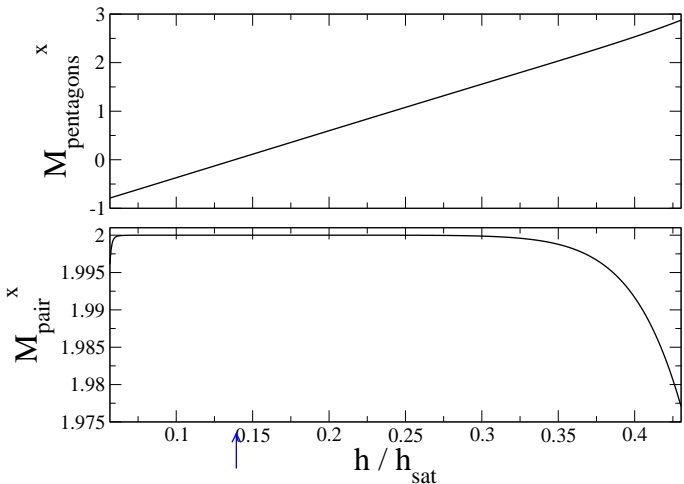}
\end{center}
\caption{Total ground-state magnetization along the field axis of the pair of spins $M_{pair}^x$ and the pentagons $M_{pentagons}^x$ shown in Fig. \ref{fig:icosahedronclusterh=1} as a function of the magnetic field over its saturation value $\frac{h}{h_{sat}}$. The (blue) solid arrow points at the location of $h=J$.
}
\label{fig:pairpentagonspinz}
\end{figure}

In the vicinity of $h=J$ the two spins related by spatial inversion point almost along the field, and their total magnetization is parallel to it. The total magnetization of the pentagons is antiparallel to the field for $h<J$ and parallel to it otherwise (Fig. \ref{fig:pairpentagonspinz}).
The icosahedron appears as four individual magnetization units, with the pentagons
forming two well-separated units whose combined magnetization along the field increases, going through the zero value.
Fig. \ref{fig:polaranglesxxicosahedron} shows that the ground-state polar angles around $h=J$ do not vary in a uniform way with the field. The same holds true for the nearest-neighbor correlation functions in Fig. \ref{fig:correlationsxxicosahedron}. This is because the field range where the two isolated spins are almost aligned with the field is associated with a ground-state degeneracy, which corresponds to a degeneracy in the polar angles and the loss of reflection symmetry along the $x$ axis. The polar angles are different in degenerate ground states, while Fig. \ref{fig:polaranglesxxicosahedron} plots only one ground-state polar angle configuration for each magnetic field, and similarly for the correlations in Fig. \ref{fig:correlationsxxicosahedron}. As in the $h=J$ case, the total energies and magnetizations of the individual pentagons are invariant in the different degenerate ground states, as are their total interaction energies with the isolated spins and with one another.

\begin{figure}[h]
\begin{center}
\includegraphics[width=3.2in,height=2.4in]{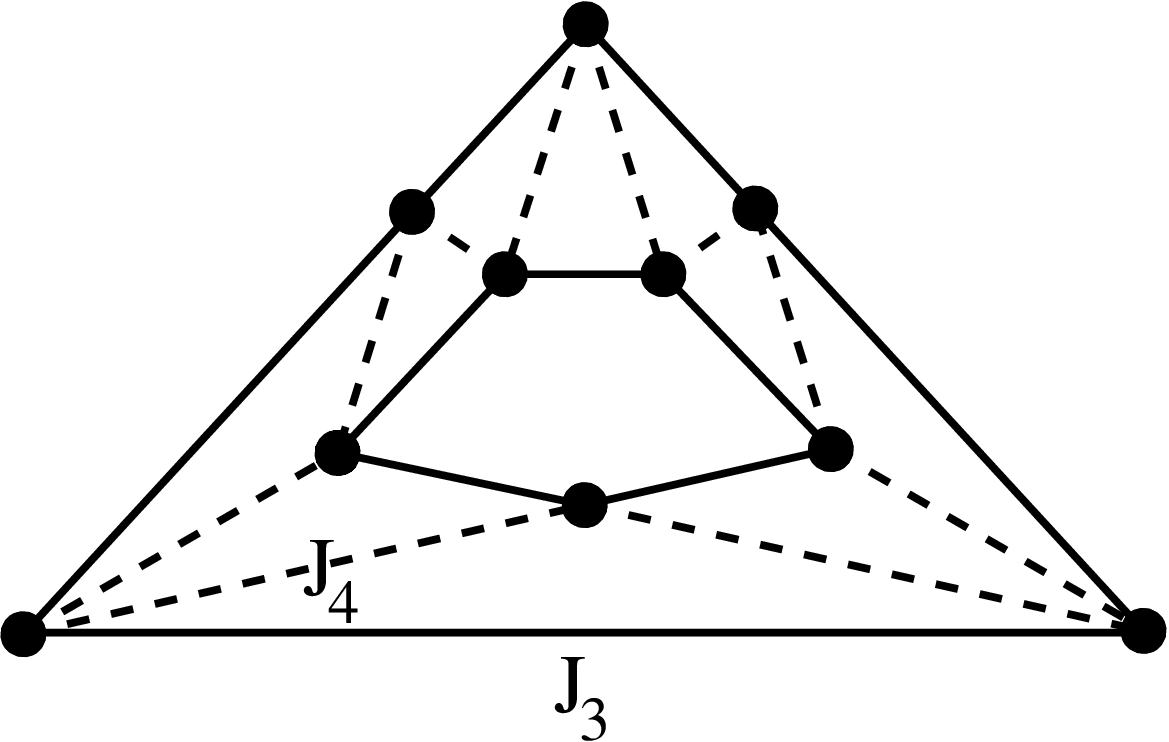}
\end{center}
\caption{Projection on a plane of the two pentagons interacting with each other. The circles are classical spins of unit magnitude and each interacts with its four nearest neighbors as shown by the connecting lines. The solid and dashed lines indicate interactions of strength $J_3$ and $J_4$ respectively. At the spatially isotropic limit $J_4=J_3$.
}
\label{fig:icosahedronminustwo22}
\end{figure}

The origin of the degeneracy can be traced to the ground state of the AXXM in a field on the 10-site cluster that is missing the two spins of the icosahedron related by spin inversion (Fig. \ref{fig:icosahedronminustwo22}). This structure is part of a triangular lattice that folds back to itself with periodic boundary conditions. Another way to look at it is as a closed chain with nearest and next-nearest-neighbor interactions, interpentagon $J_4$ and intrapentagon $J_3$ respectively. The Hamiltonian for this triangular strip with these interactions and a magnetic field $h'$ along the $x$ axis is:

\begin{eqnarray}
H & = & J_3 \sum_{<ij>'} ( s_i^x s_j^x + s_i^y s_j^y ) + J_4 \sum_{<ij>''} ( s_i^x s_j^x + s_i^y s_j^y ) \nonumber \\ & & - h' \sum_{i=1}^{N-2} s_i^x
\label{eqn:Hamiltoniantwoparts}
\end{eqnarray}

The two couplings are parametrized with the angle $\omega_{3,4}$, so that $J_3=cos\omega_{3,4}$ and $J_4=sin\omega_{3,4}$, with $0 \leq \omega_{3,4} \leq \frac{\pi}{4}$. Introducing a weak interpentagon coupling brings about the degeneracy for small magnetic fields. The ground-state polar angles are plotted for $\omega_{3,4}=\frac{\pi}{100}$ in Fig. \ref{fig:icosahedronminustwoscalePIomega=0.01PI}. For weak fields they do not follow a uniform pattern due to the degeneracy of the ground state, unlike what happens for higher fields. A finite $J_4$ introduces the triangles as structural units associated with interactions in the cluster, and the ground state of an isolated triangle is degenerate with respect to the polar angles. The ground-state energy of the AXXM in a magnetic field on a triangle can be achieved even for unequal relative angles between the spins and for different spin angles with the magnetic field, as long as the exchange and magnetic energies equal the ones of the ground state (App. \ref{appendix:XXtriangle}). The AXXM in a field ground states are also planar ground states of the AHM on a triangle in a field. The low-field degenerate ground state shares the characteristics with the one of the icosahedron in the vicinity of $h=J$. Further increasing the strength of the interpentagon coupling does not diminish the degeneracy, as shown for the polar angles in the ground state of $\omega_{3,4}=\frac{\pi}{4}$ in Fig. \ref{fig:icosahedronminustwoscalePIomega=0.25PI}. This is the spatially-isotropic interaction limit and the degeneracy is still present, having moved to higher magnetic fields.

\begin{figure}[h]
\begin{center}
\includegraphics[width=3.5in,height=2.5in]{icosahedronminustwoscalePIomega=0.01PI}
\end{center}
\caption{Low-field ground-state polar angles $\phi_i$, $i=1,\dots,N-2$ of Hamiltonian (\ref{eqn:Hamiltoniantwoparts}) as a function of the magnetic field over its saturation value $\frac{h'}{h'_{sat}}$ for the 10-site cluster and $\omega_{3,4}=\frac{\pi}{100}$.
}
\label{fig:icosahedronminustwoscalePIomega=0.01PI}
\end{figure}

\begin{figure}[h]
\begin{center}
\includegraphics[width=3.5in,height=2.5in]{icosahedronminustwoscalePIomega=0.25PI}
\end{center}
\caption{Low-field ground-state polar angles $\phi_i$, $i=1,\dots,N-2$ of Hamiltonian (\ref{eqn:Hamiltoniantwoparts}) as a function of the magnetic field over its saturation value $\frac{h'}{h'_{sat}}$ for the 10-site cluster and $\omega_{3,4}=\frac{\pi}{4}$.
}
\label{fig:icosahedronminustwoscalePIomega=0.25PI}
\end{figure}

The transition from the triangular-lattice like structure to the spatially isotropic icosahedron is accomplished by introducing the coupling to the two spatial-inversion related additional spins (Fig. \ref{fig:icosahedroncluster1}). The Hamiltonian is (compare with the AHM in a field in Ref. \cite{NPK15}):

\begin{figure}[h]
\begin{center}
\includegraphics[width=3.5in,height=2.5in]{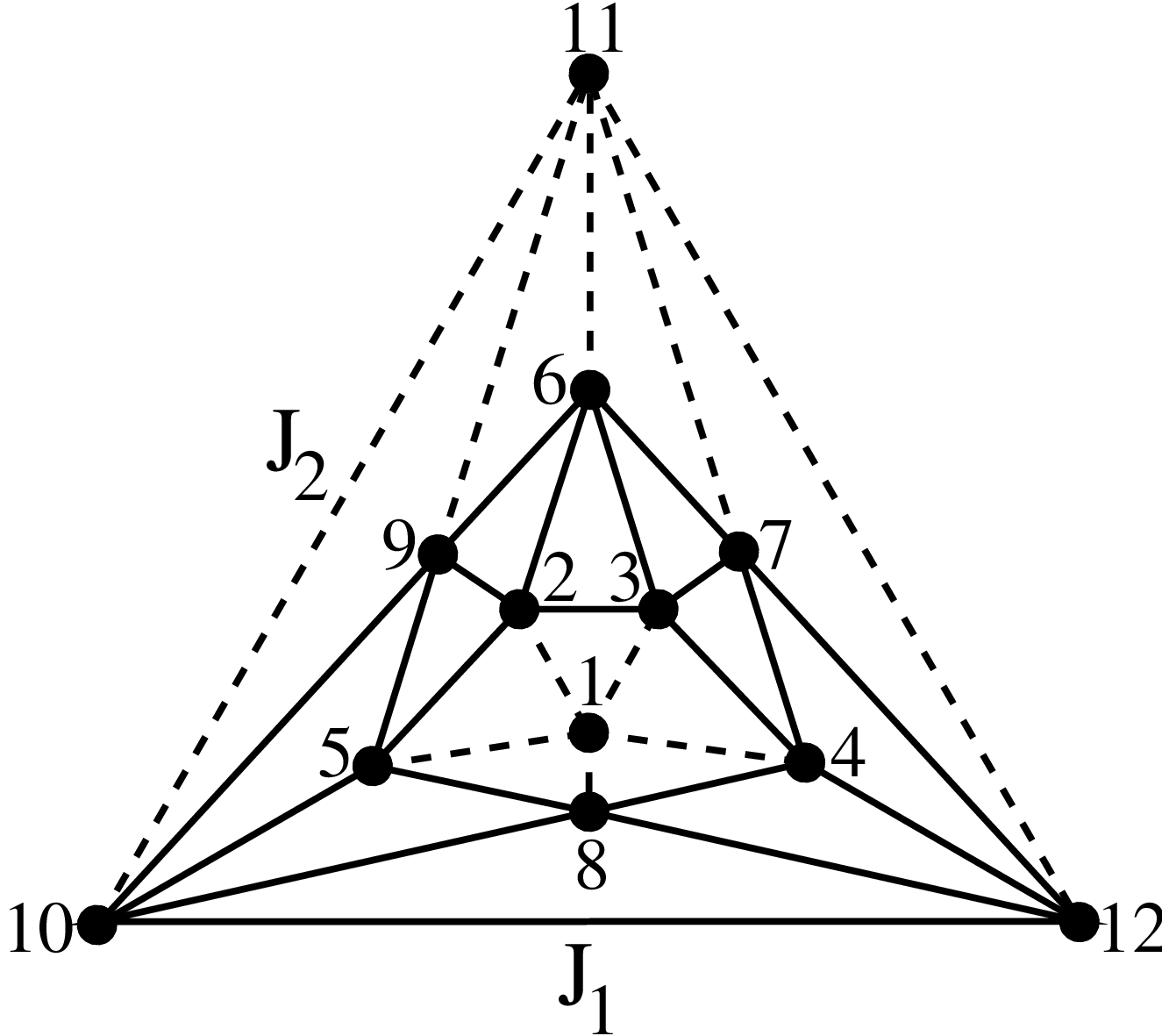}
\end{center}
\caption{A projection of the icosahedron on a plane. The circles are classical spins of unit magnitude and each interacts with its five nearest neighbors as shown by the connecting lines. The solid and dashed lines indicate interactions of strength $J_1$ and $J_2$ respectively. At the spatially isotropic limit $J_2=J_1$.
}
\label{fig:icosahedroncluster1}
\end{figure}

\begin{eqnarray}
H & = & J_1 \sum_{<ij>,i,j \neq 1,11} ( s_i^x s_j^x + s_i^y s_j^y ) + J_2 \vec{s}_1 \cdot ( \vec{s}_2 + \vec{s}_3 + \nonumber \\ & & \vec{s}_4 + \vec{s}_5 + \vec{s}_8 ) + J_2 \vec{s}_{11} \cdot ( \vec{s}_6 + \vec{s}_7 + \vec{s}_9 + \vec{s}_{10} + \vec{s}_{12} ) - \nonumber \\ & & h \sum_{i=1}^{N} s_i^x
\label{eqn:Hamiltoniantwoparts2}
\end{eqnarray}
$J_1$ is the exchange interaction between spins in the 10-site cluster, and $J_2$ the interaction between these spins and the two spatial-inversion related additional spins. The exchange interactions are here parametrized as $J_1=cos\omega_{1,2}$ and $J_2=sin\omega_{1,2}$, with $0 \leq \omega_{1,2} \leq \frac{\pi}{4}$. For $\omega_{1,2}=0$ the two spins are disconnected from the rest of the cluster and interact only with the magnetic field, while $\omega_{1,2}=\frac{\pi}{4}$ corresponds to the spatially isotropic limit of the icosahedron and Hamiltonian (\ref{eqn:model}).

Switching on the $J_2$ interaction results in the degenerate ground state window moving toward higher magnetic fields. This is shown in Figs \ref{fig:polaranglesomega=0p1PI} and \ref{fig:polaranglesomega=0p2PI} that plot the ground-state polar angles for $\omega_{1,2}=\frac{\pi}{10}$ and $\frac{\pi}{5}$. The relative strength of the triangular-type coupling increases at the expense of the pentagon-type one for higher $\omega_{1,2}$. For the spatially isotropic coupling $\omega_{1,2}=\frac{\pi}{4}$ the degenerate ground state moves to even higher fields (Fig. \ref{fig:polaranglesxxicosahedron}).

\begin{figure}[h]
\begin{center}
\includegraphics[width=3.5in,height=2.5in]{polaranglesomega=0p1PI}
\end{center}
\caption{Ground-state polar angles $\phi_i$, $i=1,\dots,N$ of Hamiltonian (\ref{eqn:Hamiltoniantwoparts2}) as a function of the magnetic field over its saturation value $\frac{h}{h_{sat}}$ for the icosahedron and $\omega_{1,2}=\frac{\pi}{10}$.
}
\label{fig:polaranglesomega=0p1PI}
\end{figure}

\begin{figure}[h]
\begin{center}
\includegraphics[width=3.5in,height=2.5in]{polaranglesomega=0p2PI}
\end{center}
\caption{Ground-state polar angles $\phi_i$, $i=1,\dots,N$ of Hamiltonian (\ref{eqn:Hamiltoniantwoparts2}) as a function of the magnetic field over its saturation value $\frac{h}{h_{sat}}$ for the icosahedron and $\omega_{1,2}=\frac{\pi}{5}$.
}
\label{fig:polaranglesomega=0p2PI}
\end{figure}

\begin{figure}[h]
\begin{center}
\includegraphics[width=3.5in,height=2.5in]{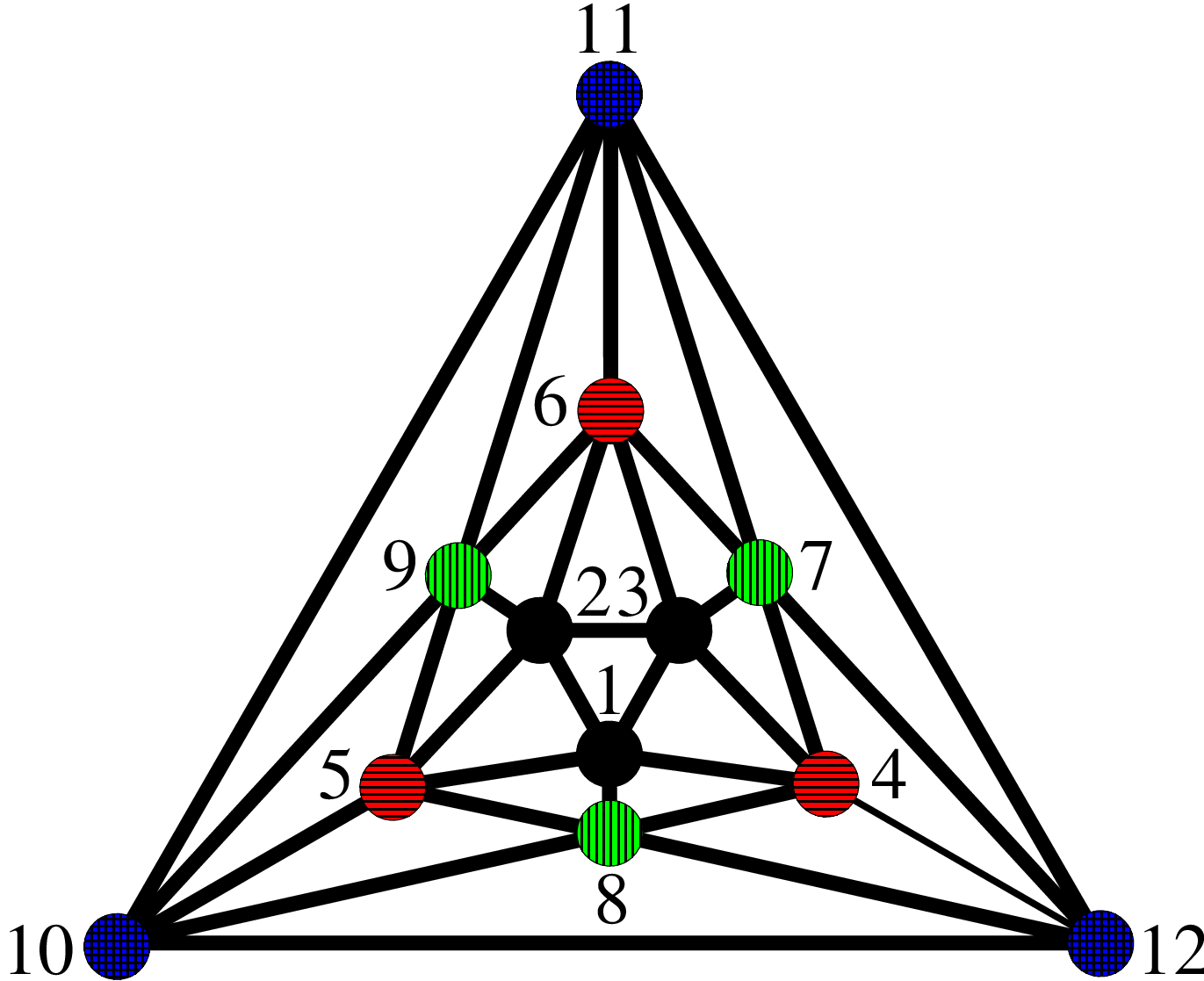}
\end{center}
\caption{Polar-angle configuration of the ground state of Hamiltonian (\ref{eqn:model}) above the second magnetization discontinuity for the icosahedron. Circles with the same pattern (and color) correspond to the same polar angle. The polar angles of spins 1 and 10 add up to $2\pi$. The polar angles of spins 4 and 7 add up to $2\pi$. 
}
\label{fig:icosahedronclusterxx2}
\end{figure}

The four-unit configuration of the ground state of Hamiltonian (\ref{eqn:model}) ceases to exist at $\frac{h}{h_{sat}}=0.4303390$, with the lowest-energy configuration the one of Fig. \ref{fig:icosahedronclusterxx1}. Figure \ref{fig:icosahedronclusterxx2} shows the higher-symmetry ground state for higher magnetic fields, entering with the second magnetization discontinuity (Fig. \ref{fig:magnetizationxxicosahedron} and Table \ref{table:magnetizationdiscontinuities}). It is described by two unique polar angles and its functional form leads to the calculation of the saturation field $h_{sat}=(5+\sqrt{5})J$ (App. \ref{appendix:saturationfield}). It is noted that nearest-neighbor correlations can be negative for fields up to $\frac{h}{h_{sat}}=0.823645$ (Fig. \ref{fig:correlationsxxicosahedron}).

\section{Connectivity and Magnetic Response}
\label{sec:connectivityandmagneticresponse}

The AHM in a field on the icosahedron has been shown to have a ground-state magnetization discontinuity that originates from the two extra spins, related by spatial inversion, being connected to the rest of the icosahedron, which resembles the triangular-lattice structure (Fig. \ref{fig:icosahedroncluster1}) \cite{NPK15}. The latter does not have any discontinuities itself. An infinitesimal coupling with the two extra spins suffices for the appearance of the discontinuity, which persists all the way up to the isotropic spatial interactions limit.

\begin{figure}[h]
\begin{center}
\includegraphics[width=3.5in,height=2.5in]{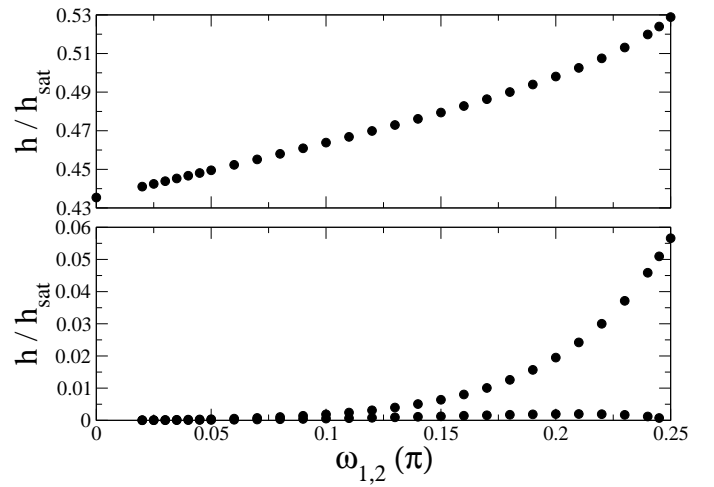}
\end{center}
\caption{Magnetic field over its saturation value $\frac{h}{h_{sat}}$ for which magnetization discontinuities occur in the ground state of Hamiltonian (\ref{eqn:Hamiltoniantwoparts2}) as a function of $\omega_{1,2}$.
}
\label{fig:disctwoextra}
\end{figure}

To investigate the role of connectivity in the appearance of the two magnetization discontinuities in the ground state of Hamiltonian (\ref{eqn:model}) the same spatial decomposition of the icosahedron in two parts is applied, which leads to Hamiltonian (\ref{eqn:Hamiltoniantwoparts2}) (Fig. \ref{fig:icosahedroncluster1}). Figure \ref{fig:disctwoextra} plots the magnetic field values for which magnetization discontinuities occur in the ground state of Hamiltonian (\ref{eqn:Hamiltoniantwoparts2}) as a function of $\omega_{1,2}$. As in the case of the AHM, an infinitesimal coupling generates low-field discontinuities and eventually one of them develops into the low-field discontinuity of Hamiltonian (\ref{eqn:model}) at the icosahedron limit, with the other disappearing exactly at that limit. However, unlike the antiferromagnetic Heisenberg model, the higher-field discontinuity is already present at $\omega_{1,2}=0$ and does not originate from the coupling of the two extra spins to the triangular-lattice like structure.

\begin{figure}[h]
\begin{center}
\includegraphics[width=3.2in,height=2.4in]{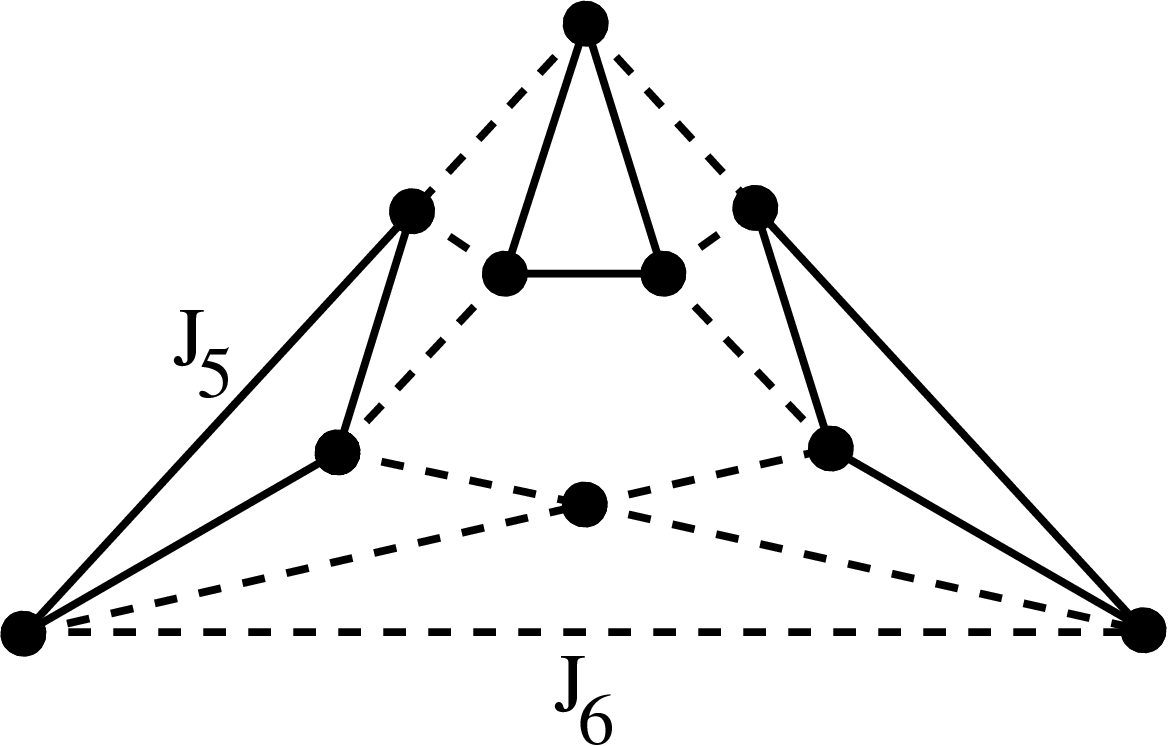}
\end{center}
\caption{Projection on a plane of the closed triangular strip consisting of three isolated triangles and an extra spin. The circles are classical spins of unit magnitude and each interacts with its four nearest neighbors as shown by the connecting lines. The solid and dashed lines indicate interactions of strength $J_5$ and $J_6$ respectively. At the spatially isotropic limit $J_6=J_5$.
}
\label{fig:icosahedroncluster2}
\end{figure}

\begin{figure}[h]
\begin{center}
\includegraphics[width=3.5in,height=2.5in]{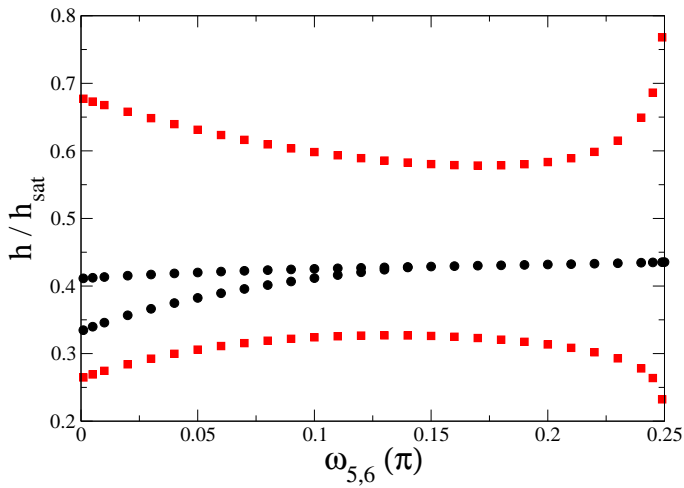}
\end{center}
\caption{Magnetic field over its saturation value $\frac{h'}{h'_{sat}}$ for which magnetization (black circles) and susceptibility (red squares) discontinuities occur in the ground state of Hamiltonian (\ref{eqn:Hamiltoniantwoparts}) with $J_3$ and $J_4$ replaced by $J_5$ and $J_6$ respectively, as a function of $\omega_{5,6}$.
}
\label{fig:discfromtrianglelimit}
\end{figure}

To trace the origin of the higher-field discontinuity the triangular-lattice like structure is further analyzed to three isolated triangles with intratriangle interactions $J_5$ and intertriangle interactions $J_6$, according to Fig. \ref{fig:icosahedroncluster2}. The exchange interactions are parametrized as $J_5=cos\omega_{5,6}$ and $J_6=sin\omega_{5,6}$, with $0 \leq \omega_{5,6} \leq \frac{\pi}{4}$. The Hamiltonian is given by Eq. (\ref{eqn:Hamiltoniantwoparts}), with $J_3$ and $J_4$ replaced by $J_5$ and $J_6$ respectively. In Fig. \ref{fig:discfromtrianglelimit} two magnetization discontinuities develop for an infinitesimal coupling that eventually merge and result in the single jump that occurs when the two exchange interactions are equal. On the other hand, the two emerging susceptibility discontinuities disappear at the spatially isotropic coupling limit.

\section{Conclusions}
\label{sec:conclusions}

In this paper two-dimensional classical spins mounted on the vertices of the icosahedron that interact according to the AXXM were considered. The frustration associated with the connectivity of the icosahedron is shown to lead to two magnetization discontinuities in an external field in the ground state. The first discontinuity brings about a lowest-energy configuration where two spins related by spatial inversion are almost aligned with the field, while the rest arrange themselves so that two pentagons form well-defined magnetization units in energy and spin. The pentagon configurations are achieved with different orientations of the individual spins, leading to ground-state degeneracy for a wide field range. Ground-state degeneracies of this type have been linked with classical spin liquids. The degeneracy is traced to the interpentagon interactions in the molecule, which introduce the triangle as a structural unit. The two magnetization discontinuities originate from the coupling of isolated triangles and then the addition of the two spins related by spatial inversion to the rest of the icosahedron.

\section{Acknowledgment}

I have been very fortunate to have met Prof. Richter, a leading theorist in frustrated magnetism, in many conferences. I very fondly remember the discussions I had with him and his enthusiasm about science. He will continue to be a source of inspiration.

\begin{appendix}

\section{Ground-State Energy Functional for $h=J$}
\label{appendix:h=J}

At $h=J$ the ground-state energy functional of Hamiltonian (\ref{eqn:model}) is (the angle $\omega$ ranges continuously from 0 to $\frac{\pi}{5}$):
\begin{eqnarray}
E & = & 10J(cos\frac{3\pi}{5}+cos\frac{4\pi}{5})+(J-h)[cos\omega+ \nonumber \\ & & cos(\frac{\pi}{5}+\omega)+cos(\frac{2\pi}{5}+\omega)+cos(\frac{3\pi}{5}+\omega)+ \nonumber \\ & & cos(\frac{4\pi}{5}+\omega)+cos(\pi+\omega)+cos(\frac{6\pi}{5}+\omega)+ \nonumber \\ & & cos(\frac{7\pi}{5}+\omega)+cos(\frac{8\pi}{5}+\omega)+cos(\frac{9\pi}{5}+\omega)] - \nonumber \\ & & 2h
\end{eqnarray}
This eventually leads to:
\begin{eqnarray}
E & = & -5\sqrt{5}J-2h
\end{eqnarray}

\section{Ground state of the AXXM in a magnetic field on a triangle}
\label{appendix:XXtriangle}

The ground state of Hamiltonian (\ref{eqn:model}) on a triangle has energy $-\frac{3+h^2}{2}$. The total magnetization along the field $M^x=h$, while perpendicular to the field the total magnetization $M^y=0$. The angles the three spins form with the magnetic field, $\phi_1$, $\phi_2$, and $\phi_3$, satisfy the equations:
\begin{eqnarray}
& & cos(\phi_1-\phi_2)-h(cos\phi_1+cos\phi_2)=-\frac{1+h^2}{2} \nonumber \\
& & cos\phi_1+cos\phi_2+cos\phi_3=h \nonumber \\
& & sin\phi_1+sin\phi_2+sin\phi_3=0
\end{eqnarray}
For a specific value of $h$ these conditions are satisfied for a degenerate manifold of the three angles.

\section{Saturation Field}
\label{appendix:saturationfield}

In the high-field ground state of Hamiltonian (\ref{eqn:model}) for the icosahedron (Fig. \ref{fig:icosahedronclusterxx2}) there are four distinct nearest-neighbor correlations. The ground-state energy functional is:

\begin{eqnarray}
\frac{E}{6}-J & = & J [ 2cos(\phi_1-\phi_2) + cos(\phi_1+\phi_2)  + cos(2\phi_2) ]  - \nonumber \\ & & h ( cos\phi_1 + cos\phi_2 )
\end{eqnarray}

Close to saturation $\phi_1 \to 0$ and $\phi_2 \to 0$ and a small-angle expansion up to second order gives:

\begin{eqnarray}
\frac{E}{6}-J & \approx & J \{ 2[1-\frac{(\phi_1-\phi_2)^2}{2}] + 1-\frac{(\phi_1+\phi_2)^2}{2}  + 1- \nonumber \\ & & \frac{(2\phi_2)^2}{2} \}  - h (1-\frac{\phi_1^2}{2}+1-\frac{\phi_2^2}{2} )
\end{eqnarray}
This eventually leads to:
\begin{eqnarray}
\frac{E}{6}-5J+2h & \approx & -\frac{1}{2}(3J-h)\phi_1^2-\frac{1}{2}(7J-h)\phi_2^2+ \nonumber \\ & & J\phi_1\phi_2 )
\end{eqnarray}

The partial derivatives with respect to $\phi_1$ and $\phi_2$ are:

\begin{eqnarray}
\frac{\partial (\frac{E}{6}-5J+2h)}{\partial \phi_1} & \approx & -(3J-h)\phi_1+J\phi_2  \nonumber \\
\frac{\partial (\frac{E}{6}-5J+2h)}{\partial \phi_2} & \approx & -(7J-h)\phi_2+J\phi_1
\end{eqnarray}

The minimum is achieved when the partial derivatives equal zero, leading to the equation:


\begin{eqnarray}
h^2-10Jh+20J^2=0
\end{eqnarray}



from which the saturation field $h_{sat}=(5+\sqrt{5})J$, identical to the value for the AHM \cite{Schroeder05}.

\end{appendix}

\bibliography{icosahedronxx}

\end{document}